\documentclass[12pt]{article}
\usepackage{epsfig}

\begin{document}
\title{Persistent Non-ergodic fluctuations in mesoscopic insulators}
\author{Rafael Rangel$\ast\star\diamond$ and Ernesto Medina $\dagger$\\
 Institut fuer Computer Anwendungen,ICA1,Univ.Stuttgart\\
 Pfaffenwaldring 27, 70569 Stuttgart,Germany$\ast$\\
Departamento deF\'\i sica, Universidad Sim\'on Bol\'\i var,\\
Apdo. 89000,Caracas 1080A, Venezuela $\star$\\ Laboratorio de
F\'\i sica Estad\'\i stica de SistemasDesordenados\\
 Centro de F\'\i sica, IVIC,Apdo.21827\\ Caracas
1020A,Venezuela$\dagger$}

\maketitle

\begin{abstract}
We give a detailed  and rigorous  picture of the mesoscopic
conductance fluctuations in the deep insulating regime (DIR)
within the Nguyen, Spivak and Shklovskii model including
spin-orbit coupling (SO). Without SO, we find that
fluctuations of the log-conductance are persistent above a saturation field $%
B_s$, where one has that the log-conductance is approximately
 a stationary random process. In contrast, in  the SO case the
saturation field $B_s$ is negligible and the stationarity  is
well realized. We find  non-vanishing  disorder fluctuations of
the field average of the log-conductance  as a quantitative
measure of the lack of ergodicity in the mean square sense. To
this fact, a weak decaying behavior of the correlation function,
even weaker in the case of SO is established  on the relevant
field  scale of the model, one flux quantum per plaquette. This
finding corroborate the behavior of the fluctuations of the field
average of the log-conductance and permit us to invoke Slutski's
theorem to conclude that the whole stochastic process defined by
the log-conductance is non-ergodic  in the mean square sense in
both cases. As a consequence the commonly used criterion to test
the ergodicity based on the equivalence  of  the variance in
disorder
 and  the variance in the field   is not fulfilled. Using
the replica approach, we derive the weak localization analogs of
the `cooperon and diffuson which permits us to analyze in
qualitative form  the decaying behavior of the correlation
function.  Our predictions agree qualitatively  and
semi-quantitative with experiments in the DIR.
\end{abstract}
\newpage

\section{Introduction}
\label{sec:intro}
 The nature  of Fluctuations in both the metallic state
\cite{LeeStoneFukuyama},\cite{JRammer}, and in disordered
insulators \cite{LadieuBouchaud}, \cite
{LadieuMaillySanquer,Koslov}, has been a matter of interest for
both theoretical and experimental studies. Whereas in the metallic
regime the basic aspects of fluctuations  have been elucidated, in
the regime of hopping transport the nature of fluctuations is
still an open field.  The deep insulating regime, DIR, where
transport occurs  via variable range hopping (VRH), is defined as
the regime where the localization length is the smallest scale
compared to the elastic mean free path and hopping lengths, i.e.,
$\xi <{\ell }<t$ respectively\cite{ShklovskiiSpivak}. Coherence
effects are possible in this regime because phase breaking events
occur at the hopping length\cite{NSS}, which is larger than ${\ell
}$. Important signatures of quantum interference in disordered
insulators are the classic magneto-fingerprints, or reproducible
fluctuations in the conductance with magnetic field, and  a low
field positive magneto-conductance.

 An important property of mesoscopic conductance fluctuations in
 the metallic phase is their
ergodicity. At  the mesoscopic level,  the sample size is less
than thermal diffusion length or the dephasing length, whatever is
shorter, such that sample to sample fluctuations are visible and
the system does not  self-average. Although it was not rigorously
proven, the ergodic hypothesis was meant as the ability of the
magnetic field ( or energy) to induce conductance fluctuations
equivalent to sample to sample fluctuations (Lee-Stone criterion)
\cite{LeeStoneFukuyama}. In contrast, experimental results show
that log-conductance mesoscopic fluctuations in the DIR without
spin-orbit scattering are not ergodic  in the Lee-Stone sense\cite
{LadieuMaillySanquer,Koslov,LaikoOSIP}, i.e., the variance over
samples is larger than the variance over field. Such samples
involve hopping lengths that are, at most, 6 to 10 times the
localization length. Precise measurements  of Ladieu et
al\cite{LadieuMaillySanquer} and  Orlov et al\cite{Koslov} have
shown that a) field fluctuations do not decorrelate disorder
fluctuations,  b) field fluctuations do not change the identity of
the hop, c) the field average of the variance over the samples is
larger than the sample average of the variance over the field and
d) there exist a decorrelation field $B_{c}$ defined by the field
correlation function, which defines a equivalent  new sample.

The question of the ergodic nature of  fluctuations with  and
without SO has not been addressed, to our knowledge, from the
theoretical side. Our plan of this work is first to address the
problem of fluctuations and the question of ergodicity in DIR
within the NSS model. We undertake this task through the
verification of concepts concerning ergodicity, which we first
define with mathematical rigor in the next  mostly technical
section. Then in section \ref{sec:NSSmodel} we explain the NSS
model and define the random processes we want to analyze. In
section\ref{sec:Fluctuations} we carry out the program described
in section \ref{sec:ergodicconcepts} and verify the non-ergodic
behavior of fluctuations. In section \ref{sec:corrCoopDiff} we
define the main theoretical objects of this work, the cooperon and
diffuson analogs of weak localization theory, with the help of
which, we can explain  to some degree the decaying behavior of the
correlation function. Finally we conclude by discussing and
comparing our results with experiments.

\section{ Ergodicity of Transport Fluctuations}
\label{sec:ergodicconcepts}

 In the preceding section one we introduced the question of ergodicity of fluctuations.
 Here
 we introduce the mathematical  concepts that will permit us  to establish the ergodic nature of the fluctuations.

  Given a
physical quantity $F({\cal H},B)$ depending on the disordered
Hamiltonian ${\cal H}$ and magnetic field $B$,  we denote by
$\overline{F({\cal H},B)}$ the sample to sample  average,
or disorder average,  and by   $\langle F({\cal H}%
,B)\rangle =\Delta B^{-1}\int_{B_{i}}^{B_{f}}dB~F({\cal H},B)$ the
field average for a given sample or disorder realization. In order
to be able to estimate the sample average from the field average
of a given sample the following conditions must be satisfied:  a) $%
\lim_{B_{f}\rightarrow \infty }$ $\sigma _{mss}(B_{f})=\overline{[\overline{%
F({\cal H},B)}-\langle F({\cal H},B)\rangle ]^{2}}\rightarrow 0$
and    b) $\overline{F({\cal H},B)}=\overline{\langle F({\cal
H},B)\rangle }$. The verification of both conditions is known as
{\it ergodicity in the mean square sense} (mss), or the random
function ${F({\cal H},B)}$ is
said to be ergodic in the mean-square limit\cite{AMYaglom}. The condition $\overline{F({\cal H},B)}=%
\overline{\langle F({\cal H},B)\rangle }$ is a measure of global
stationarity. which  means that these averages are independent of
$B$. One can cast  conditions a) and b)  into a single statement
on the disorder fluctuations of the
field average,i.e.,  $%
\lim_{B_{f}\rightarrow \infty }\sigma _{mss}(B_{f})=\lim_{B_{f}\rightarrow
\infty }{\rm Var}_{d}(\langle F({\cal H},B)\rangle)=\overline{(\langle F({\cal H},B)\rangle-\overline{%
\langle F({\cal H},B)\rangle})^{2}}\rightarrow 0$, where  $\ {\rm
Var}_{d}$ means variance over disorder\cite{Pandey}. This property
implies  that for one realization of disorder there are enough
equivalent samples within  the magnetic scale, such that the
average in the field, with regard to disorder, does not depend in
statistical sense  on the particular realization. This means that
$ \overline{\langle F({\cal H},B)\rangle }$$\approx {N {\langle
F({\cal H},B)\rangle} } $, where $N$ is the number of
realizations, and a sharp distribution of $ \langle F({\cal
H},B)\rangle $ over disorder holds.

One could also ask for the possibility of making estimates of $\ {\rm Var}%
_{d}(F({\cal H},B))$ from $\ {\rm Var}_{B}(F({\cal H},B))$, ( here $\ {\rm %
Var}_{B}(F({\cal H},B))$ means the variance over the field B of  $F(%
{\cal H},B)$), or more generally, try to make an estimate of
another function of the basic process $F({\cal H},B),\ g(F({\cal
H},B))$.
The necessary and sufficient conditions such that one can  estimate $%
\overline{g(F({\cal H},B))}$ from one realization of disorder
 with the field  average $\langle g(F({\cal H}%
,B))\rangle$( the so called  law of strong numbers), is given by
Slutski's theorem. One can write: $\sigma
_{mss}(B_{f})$$=$$\lim_{B_{f}\rightarrow\infty }\frac{2}{\Delta
B^{2}}$ $\int_{B_{i}}^{B_{f}}dB(B_{f}-B)C(~g(F({\cal H},B))) $,
with:
\begin{equation}
C(g(F({\cal H},B,\Delta B)))=\overline{\Delta [g(F({\cal
H},B+\Delta B))\times \Delta [g(F({\cal H},B))}
 \label{defcorrelationfunction}
\end{equation}
where  $C(g(F({\cal H},B,\Delta B)))$ is the correlation function for   $g(F({\cal H%
},B,\Delta B))$ and $\Delta [g(F({\cal H},B+\Delta B))= {[g(F({\cal H},B+\Delta B))-\overline{%
\ g(F({\cal H},B+\Delta B))}\ ]}$.  One can easily realize that a
strong  decaying behavior of the correlation function with
correlation lenght $\tau\ll B_{f}$ will be a sufficient condition
for ergodicity in the mss, i.e.  $\sigma _{mss}(B_{f})\Rightarrow
0 $. In fact, it is also a necessary condition.

Usually, for the application of the theorem   stationarity in the wide sense  of $F({\cal H}%
,B)$ can be assumed ,i.e., $\overline{(F({\cal H},B))}$ does not depend on $B$ and  $C(g(F({\cal H},B,\Delta B)))\ $%
should depends only on $\Delta B$\cite {Ergodicitysubleties}. In
this work we are interested in
testing the ergodicity in two cases: $g(X)=X $ for the average and $%
g(X)=X^{2}-{\overline{X}^{2}} $for the variance. The first case
corresponds to the usual meaning of ergodicity in statistical
mechanics. The second case, usually named the Lee-Stone criterion,
refers to the equivalence of sample to sample fluctuations and
magnetic field fluctuations.
\cite{LeeStoneFukuyama}\cite{AMYaglom}\cite {FengPichard}.
\section{The NSS Model}
\label{sec:NSSmodel}

  We examine  now the fluctuations in the DIR  in
two and three dimensions and define  the random process $F({\cal
H},B)$. This  is obtained from the Nguyen, Spivak and
Shklovskii\cite{ShklovskiiSpivak,NSS} model (NSS). The NSS model's
crucial insight is that coherence is maintained within a Mott
hopping length, where the conductance is a sum of coherent {\it
forward directed Feynman paths} which interfere which each other.
The NSS model describes the quantum behavior of the critical
(bottleneck) hop in the Miller-Abrahams network\cite
{EfrosShklovskii}. The existence of many  randomly oriented
critical hops tend to average the macroscopic conductance,
eliminating fluctuations\cite{Fowler}. Here, we focus on the low
temperature regime where critical hops do not trivially
self average\cite{LaikoOSIP}\cite{SanquerRev}, i.e., the percolation correlation length $%
\xi _{p}$ is such that $\xi _{p}=\xi (T_{o}/T)^{(\nu
+1)/(D+1)}\sim L$, where $\nu $ is the percolation correlation
length exponent, $D$ is the spatial dimension and $T_{o}$ a
disorder parameter. This is the mesoscopic regime \cite{LaikoOSIP}.
 We will first find the fields above which the process $F({\cal H%
},B)$ can be considered stationary and test the ergodicity
criterion in the mean square limit. We then calculate the
correlations functions and their decaying behavior. Invoking
Slutski's theorem, we confirm non-ergodic behavior
as established by the finiteness of the mean square criterion, ($%
\lim_{B\rightarrow \infty }{\rm Var}_{d}(\langle F({\cal
H},B)\rangle)$ $\neq 0$). We then proceed to test the ergodicity
of fluctuations. Non-ergodicity of fluctuations is implied by the
 weakly decaying behavior of the correlation function. We find this
consistently. Here, we find good quantitative agreement with the
measurements of Orlov et al \cite{Koslov}. We derive  the cooperon
and diffuson analogs of the weak localization theory  for DIR with
the help of which the non ergodic behavior can be explained.
Furthermore , the predictions for the case with SO are made.

In the two dimensional NSS model, impurities are placed on the
sites of a lattice of main diagonal length $t$ (the hopping
length, $t=\xi (T_{o}/T)^{1/(D+1)},$ Mott's law). We apply a
magnetic field $B$, perpendicular to the plane, changing only the
phases of the electron paths. The overall tunneling amplitude is
computed by summing all forward directed paths between two
diagonally opposed points, each contributing an appropriate
quantum mechanical complex $2\times  2$ matrix weight given by the
Hamiltonian:
\begin{equation}
{\cal H}=\sum_{i}\epsilon _{i}a_{i,\sigma }^{\dagger }a_{i,\sigma
}+\sum_{<ij>_{{\sigma },{\sigma }\prime }}V_{ij,{\sigma },{\sigma
}\prime }a_{i,\sigma }^{\dagger }a_{j,{\sigma }\prime },
\label{Hspinor}
\end{equation}
where $\epsilon _{i}$ is the site energy, and $V_{ij,{\sigma },{\sigma }%
\prime }$ represents the nearest neighbor couplings or transfer
terms  which includes a randomly chosen SU(2) matrix describing a
spin rotation due to strong SO scattering. Within the NSS model,
we choose site energies to be $\epsilon _{i}=\pm W$
with equal probability \cite{NSS} \cite{MedinaRev}%
. Without SO the coupling terms are diagonal in spin space
$V_{ij}=V,$ and the Green's function between the initial and final
site is given by
\begin{equation}
\langle i|G(E)|f\rangle =\left( {\frac{V}{W}}\right)
^{t}J(B,t);~~J(B,t)=\sum_{\Gamma ^{\prime }}^{\scriptstyle %
directed}[\prod_{i_{\Gamma ^{\prime }}}\eta _{i_{\Gamma ^{\prime
}}}e^{i\phi_{i_{\Gamma ^{\prime}}}}], \label{eNSS}
\end{equation}
where $\phi_{i_{\Gamma ^{\prime}}}$ is the phase gained through
path $i_{\Gamma ^{\prime}}$ due to  the magnetic vector potential,
$\Gamma ^{\prime }$ represents all directed paths that go from $i$
to $f$ through the lattice and $\eta _{i}={\rm sign}\left(
\epsilon _{i}\right) =\pm 1$\cite{HansHerman}.  In the presence of
spinorbit scattering
 the Green's function is the $2 \times 2$  matrix:
\begin{equation}
J(B,t)=\sum_{\Gamma ^{\prime }}^{\scriptstyle
directed}[\prod_{i_{\Gamma ^{\prime }}}\eta _{i_{\Gamma ^{\prime
}}}][\prod_{i_{\Gamma ^{\prime }}}U_{i_{\Gamma ^{\prime
}}}]e^{\sum{i\phi_{i_{\Gamma^{\prime}}}}}\ ,\label{eNSSSO}
\end{equation}
The Green's function consist of sum of terms, one for each path,
each being a product of random numbers and random SU(2) matrices,
and a deterministic disorder independent  phase factor from the
magnetic vector potential\cite{ASchmid}. The complex function $%
J^{2}(B,t)$( a complex matrix function in the presence of SO)
contains the interference information including correlations due
to crossing of paths, and the factor $(V/W)^{t}$ is the leading
contribution to the exponential decay of the localized
wavefunction. We use the transfer matrix approach in order to
compute $J(B,t)$, exactly, for each realization of
disorder\cite{MedinaRev}. Our random processes in question
represent the log-conductance. They  are $F({\cal H},B)=\ln
(J^{\dagger}(B,t)J(B,t))\ $ and  with SO $F({\cal H},B)=\ln
(I(B,t))\ ,\ $where $I(B,t)=1/2Tr(J^{\dagger }(B,t)J(B,t))\
$\cite{Redner}. In his work we measure the magnetic field $B$ or
changes in magnetic field $\Delta B$ in flux units
$\phi_{o}/\ell^{2}$, where $\phi_{o}$ is the flux quantum.

\section{Fluctuations and Ergodicity }
\label{sec:Fluctuations}

   Figure \ref{fig1}  shows typical fluctuations of the
log-conductance  as a function of the sample and the magnetic
field. Without SO , the figure clearly shows that the average of
$\ln (J(B,t))$ dominates the fluctuations, i.e., the average
behavior is visible for a single sample.  As the average $\ln
(J(B,t))$ first increases  proportional to $B$, crossing over to a slower growth as $%
B^{1/2}$ dictated by the magnetic length $\Delta B < B_c=\pi
c\hbar/(\xi^{1/2}e~ t^{3/2})$ \cite{MedinaRev,RRangel2,Zhao}, the
process is not stationary. However, in the latter regime of slow
growth, one finds a field above which the process can be
considered as essentially stationary  in the same fashion as in
the metallic regime. In this regime, while the log-conductance
tends to saturate, the fluctuations
persist as in mesoscopic fluctuation theory in metals\cite{LeeStoneFukuyama}%
. Furthermore, we note that the average behavior is periodic in
half the flux quantum $\phi _{o}$ per $\ell ^{2}$ (only one half
of a period is shown). This periodicity reveals an average field
coupling to $2B$\cite{NSS} which has been demonstrated
theoretically\cite{RRangel1}. In three dimensions, the
fluctuations are appreciably larger than the average behavior.
Once more, persistent fluctuations beyond the average conductance
saturation field are observed. The existence of such persistent
fluctuations were first surmised by Sivan et al\cite{Fowler,rSEI}
and Zhao et al\cite{Zhao}. With SO , there is no  tendency to
build an average, and there are in general soft changes in the
fluctuations in marked contrast with the sharp changes without SO.
This peculiarity will be explained later. One marked feature of
figure 1(with and without SO), are  the fact that disorder
fluctuations do not decorrelate the field fluctuations, which  is
the reason of the  remarkable similarity with fig.2 of the work of
\cite{Koslov}. The key point in both figures is that the the
fluctuations do no decorrelate at the  scale of fields shown in
the figures, suggesting   non-ergodic behavior.

\smallskip

Following  the concepts of  section two, we analyze the ergodicity
of the log-conductance fluctuations more carefully. To achieve
this we have to check two points: first if stationarity is
reasonably fulfilled and second, we have to verify  the ergodicity
condition and  the decaying behavior of the
correlation function. Fig.\ref{fig2}a  shows the quantities  ${\rm Var}_{d}(\langle F({\cal H}%
,B)\rangle)\ $(stars) and $\overline{[\overline{F({\cal H},B)}-\langle F({\cal H}%
,B)\rangle ]^{2}}$( diamonds) in the absence of  SO ($F({\cal
H},B)=\ln|J(B,t)|^{2}$).
 The field averaging interval is  $\Delta B=[
B_{i},B_{f}]$. In this figure as  the field $%
B_{i}$ is increased  from zero to 0.09 in $[\phi_{0}/ \ell^{2}]$
units, the diamonds move downward  overlapping the stars sooner
for smaller t. Therefore, the field above which the process can be
considered quasi stationary increases with the hopping length t
such that both quantities tend to coincide for larger  $B_{i}\ $.
This feature is of importance for calculating other quantities, as
any question on ergodicity presumes  at least quasi-stationarity.
In the case of SO ($F({\cal H},B)=\ln|I(B,t)|^{2}$),
Fig.\ref{fig2}b shows effective  stationarity of the
process $\ln (I(B,t)),\ $ essentially independent of  $B_{i}$ and $t$. ${\rm Var}%
_{d}(\langle F({\cal H},B)\rangle)$ does not tend to zero with
increasing $B_{f}$; on the contrary, it increases with a power of
t\cite{RRangel1}, which is an indicator of non-ergodic behavior as
there is no self-averaging as one   increases the hopping length
t( see Kramer and Mackinnon \cite{JRammer}). The last result
establishes non-ergodic behavior in the mean square sense with and
without SO. To further substantiate  this result we calculate the
correlation functions. Fig. \ref{fig3}a shows the correlation
function for three values of  the hopping length $t$ and some
values for $B_{i}$. One sees a very weakly decaying behavior on
the physical field scale( $\phi_{o}/\ell^{2}$)  and a tendency to
decay faster for bigger $B_{i}\ $, indicating that when the
process becomes quasi stationary there is a tendency to a faster
decaying correlation function. Fig. \ref{fig3}b with SO shows that
the correlations function depends essentially only on $\Delta B,$
and the decaying is even weaker with a functional form depending
on t. Now, the basic argument against ergodicity is that the
decaying  behavior of the correlation functions is such  that it
is not possible to construct enough ensembles from the field
fluctuations data and therefore non-ergodic behavior is
established. To illustrate
this point, we write $\sigma _{mss}(B_{f})=\frac{2}{\Delta B^{2}}%
\int_{B_{i}}^{B_{f}}dB[(B_{f}-B)C(~F({\cal H},B))]$. In order to
have enough
emsembles in the field scale within the validity of the model ($B_{f}\leq $ $%
\phi _{o}/\ell ^{2}=1$ in our units) , there should be a
decorrelation field $B_{c}$ $<<1,$ such that large  number of
samples could be defined, so that one would get $\sigma
_{mss}(B_{f})\approx 0$-could be satisfied. This, on the other
hand, implies that the condition for ergodicity of the variance,
which in general requires a stronger decaying behavior of the
correlation function for  process $F({\cal H},B)\ $ than the
condition required for
ergodicity in the mean square sense, is not at all fulfilled%
\cite{AMYaglom}. Therefore, one should expect that the  Lee and
Stone criterion of ergodicity on the relative magnitude of the
field and sample fluctuations is not realized.  To check this
point one has to compare the magnitude of the variance in field
and sample to sample fluctuations. The idea of further averaging
over disorder and over the field, respectively, is the same as in
statistical mechanics using different initial conditions to
improve the statistics. Figure \ref{fig4} shows the averages
\begin{equation}
\langle {\rm Var}_{d}(\ln |J(B,t|)\rangle=\langle \overline{( \ln |J(B,t|-\overline{\ln |J(B,t|}%
)^{2}} \rangle,  \label{vardis}
\end{equation}

\begin{equation}
{\overline{{\rm Var}_{B}(\ln |J(B,t|)}}=\overline{\langle (\ln
|J(B,t)|-\langle \ln |J(B,t)|\rangle )^{2}\rangle }.
\label{varfield}
\end{equation}
\smallskip
Without SO we find the characteristic $B_{i}\ $dependence shown in
fig.\ref{fig2} shows up as  a crossing
 of both types of averages. There is again a clear tendency  to saturate as $B_{i}\ $%
increases whereas the difference of the saturated averages widens
with increasing t. This tendency is clearly seen in the case with
SO where there is a much weaker  dependence on $B_{i}\ $ as
expected from fig.\ref{fig2} and  the crossing observed in the
previous case is absent.

\section{The Correlation function: the Cooperon and Diffuson}
\label{sec:corrCoopDiff}
 In this section we develop the concepts of  the cooperon and diffuson
 in the context of strong localization. These objects  are used to explain qualitatively and semi-quantitatively
non-ergodic behavior  in the mean square sense and the relative
magnitude of the field and sample fluctuations found in
experiments. For this purpose, the following  relation  is
straightforwardly derived :\smallskip
\begin{equation}
{\rm Var}_{d}[F(B+\Delta B,t)+F(B,t)]={\rm Var}_{d}[F(B+\Delta B,t)]+{\rm Var}%
_{d}[F(B,t)]+2C(F({\cal H},B,\Delta B,t))  \label{nonlinear}
\end{equation}\smallskip

This relation is valid for both processes $\ln J(B,t)$ and $\ln
I(B)$\cite{explanationnonlinear}. For the sake of clarity, the
composite process inside the brackets on the left side of
eqn.\ref{nonlinear} is denoted by the process $P(B,\Delta
B,t)=[\ln| J(B+\Delta B)|^{2}+\ln|J(B)|^{2}]\ $and with SO by
$P_{spinor}(B,\Delta B,t)=[\ln I(B+\Delta B)+\ln I(B)]$, such that
the left hand side of eqn. \ref{nonlinear} is in
each case  ${\rm Var}_{d}[P(B,\Delta B,t)]$ and   ${\rm Var}_{d}[P_{spinor}(B,%
\Delta B,t)]$, respectively. These  two functions are shown in
fig. \ref{fig5}  as a function of $%
\Delta B\ $ in  $\phi_{0}/\ell ^{2}\ $ units  and three values of
t $(30,100,300)$ from the bottom to the top, respectively). For
$B_{i}\ $ shown in the figure and the scales of $\Delta B\ $used,
three  things deserve explanations: first for large enough $\Delta
B$, the above mentioned functions show  almost no dependence  on
$B_{i}\ $and $\Delta
B$ at fixed t; secondly,  the ratio ${\rm Var}%
_{d}[P(B,\Delta B,t)]/$ ${\rm Var}_{d}[P_{spinor}(B,\Delta B,t)]\ $ is
around two, for big enough t and  $%
\Delta B$  which is a landmark of the symmetry changing from
unitary to simpletic;  Thirdly, one observes a very rapid decaying
on the dependence on $B$, which  can be traced to the saturation
behavior of the cooperon as we will see.

 Recall that on the
metallic side, an analogous behavior has been described which
identifies two fundamental contributions to the field effect: the
cooperon and the diffuson\cite{LeeStoneFukuyama}\cite{JRammer}.
These contributions can be distinguished by the way they enclose
the magnetic flux; while the cooperon is sensitive to ($2B+\Delta
B)$, the diffuson only responds to field changes $\Delta B$. In
the insulating regime, a mechanism similar to the cooperon which
saturates is associated with a positive magneto-conductance (MC).
This  has been observed as a general effect\cite{JRammer}\cite
{LaikoOSIP,Sarachik,Koslov,FaranOvadyahu}.  A semi quantitative
explanation for the behavior of the functions   ${\rm
Var}_{d}[P(B,\Delta B,t)]\ $and ${\rm Var}_{d}[P_{spinor}(B,\Delta
B,t)]$  and their ratio can be found with the help of the cooperon
and diffuson analogs. To achieve this goal, we consider  the
moments of process of $P(B,\Delta B,t)$ and $P_{spinor}(B,\Delta
B,t)$( further below it will become clear why). In  the former
case they are given by $\overline{[J^{*}(B+\Delta B)J(B+\Delta
B)J^{*}(B)J(B)]^{n}}$. Recall from equations \ref{eNSS} and
\ref{eNSSSO} that this product can be visualized as  a set of n
paths, each one defined by the respective term in the product.
These paths eventually intersect each other at some values of
their length. In order to have nonzero contributions after
disorder average, the paths must pair up, as a consequence of th
chosen distribution of the energies \cite{Kardar}. Neutral paths
(field independent) are formed by pairing $J^{*}$ and $J$ at the
same field (phase cancels). On the other
hand, charged paths (field sensible) are formed by pairing either $%
J^{*}(B+\Delta B)$ and $J(B)$ or $J^{*}(B+\Delta B)$ and
$J^{*}(B)$. In the absence of paired path intersections, self
interference kills charged paths (their contribution decays
exponentially fast). Nevertheless, if intersections are
considered, one can have path exchanges for short distances,
yielding a magnetic field coupling which is analogous to the
magneto-conductance, the source of the initial decaying of the
correlation function. There are three possible diagrams at a
paired path crossing, two of which are depicted in fig.7. Without
SO the spin indexes can be ignored, one obtains \cite{RRangel1}:
a) one partner from $J^{*}(B+\Delta B)$ pairs with one from
$J^{*}(B)$ while one from $J(B+\Delta B)$ and one from $J(B)$
follow a different path. Such a combination encloses $(2B+\Delta
B)$ and is
therefore called {\it cooperon}-like. b) One partner is taken from $%
J^{*}(B+\Delta B)$ and the other from $J(B)$ on the same path,
while one from $J(B+\Delta B)$ and $J^{*}(B)$ follow another. Such
a combination encloses only $\Delta B$ and is called {\it
diffuson}-like. Finally, one can have combination c) where one
partner comes from $J^{*}(B+\Delta B)$ and the other from
$J(B+\Delta B)$, leaving $J^{*}(B)$ and $J(B)$ to pair up. The
latter combination is called uncharged and encloses no field. Note
that all previous cases satisfy overall neutrality so that the
contributions are real as expected. The contribution of the
replica cooperon and diffuson are the same at zero field and there
is an additional contribution from the uncharged diagram. Further
progress is achieved using the replica argument.

The replica-moment argument \cite{Kardar}, maps the $n$-th moment
problem onto the problem of $2n$ bosons with contact interaction.
These interactions renormalize due to the diagrams above, making
path interactions field dependent. The $2n$ boson system can be
solved using the Bethe ansatz and has ground state energy
$\epsilon _{0}=\ln 4^{n}+\rho (B,\Delta B)n(n^{2}-1) $, where
$\rho (B,\Delta B)$ is a function of $B,\Delta B$, such that one
has $\overline{[J^{*}(B+\Delta B)J(B+\Delta
B)J^{*}(B)J(B)]^{n}}=$$A(n,B,\Delta B)\exp (ln4^{ n}+\rho
(B,\Delta B)n(n^{2}-1)t $ $)$ valid at fixed $n$ asymptotically
for $t\rightarrow \infty $. On the other hand, the $n$-th moment
can be expressed as a
cumulant expansion valid at fixed $t$ asymptotically for $n\rightarrow 0$, $%
\overline{[J^{*}(B+\Delta B)J(B+\Delta
B)J^{*}(B)J(B)]^{n}}$ $=\exp \{\sum \frac{n^{i}}{i!}%
C_{i}[P(B,\Delta B,t]\}$, where $C_{i}[P(B,\Delta B,t]$ are the
cumulants of process P . The subtleties concerning both limits
have  been discussed by Kardar \cite{Kardar}, who finds a
nonextensive correction subleading term proportional to $t^{2/3}$.
We therefore obtain :

\begin{equation}
{\rm Var}_{d}[P(B,\Delta B,t)]=(\rho _{coop}(2B+\Delta B)+\rho
_{diff}(\Delta B))t^{2/3}+\ln A(B,\Delta B)  \label{cooperondiff}
\end{equation}
Here $\rho (B,\Delta B)=$ $\rho _{coop}(2B+\Delta B)+\rho _{diff}(\Delta B)$%
, we have separated the path interaction in terms of the cooperon
and diffuson contributions.  We check numerically this  important
prediction and fig.6 shows the expected scaling with an exponent
near $\frac{2}{3}$. Beyond the saturation field of the average
log-conductance, the cooperon term on the right hand side
saturates (the same for the variance on the right of eqn.
\ref{nonlinear} which depend on $B$) and the correlation function
only depends on $\Delta B$. This behavior is summarized in
Fig\ref{fig5}.

 In the case of SO, it can be shown\cite{MedinaRev},
that the only non-zero paired averages are
$\overline{U_{\alpha\beta}U_{\alpha\beta}^{*}}=\frac{1}{2}$,
$\overline{U_{\uparrow\uparrow}U_{\downarrow\downarrow}^{*}}=\frac{1}{2}$,
$\overline{U_{\uparrow\downarrow}U_{\downarrow\uparrow}^{*}}=-\frac{1}{2}$
, thus SO averaging brings a factor of $(\frac{1}{2})^{2}$ and
forces the neutral paths to have parallel spins while the spin of
the two partners of charged paths must be antiparallel. As a
consequence, one finds the cooperon diagrams cancel in pairs as
concluded  for the case of the  magneto-
conductance\cite{MedinaRev}(fig.7) such that no exponential
corrections to the conductance occurs  due to the cooperon. For
the diffuson there are only two combinations possible for incoming
spin
indexes(all up and all down),  such that in this case ${\rm Var}%
_{d}[P_{spinor}(B,\Delta B,t)]=$$(\rho_{diff} ^{spinor}(\Delta B)
)t^{2/3}$ $+\ln A(B,\Delta B)$. This nice result explains why
there are fluctuations with SO, even if there are not exponential
corrections to the magneto-conductance. We find again numerically
the predicted  $t^{2/3}$  scaling (see fig.6). For large $\Delta
B$ and t one finds the ratio  of the variances approaches $\approx
\rho _{diff}(\Delta B))/\rho_{diff} ^{spinor}(\Delta B)= 2$ (i.e.,
1/(2 times $(1/2)^{2}$)), in agreement with the numerical results.
 Now
from eqn. \ref {nonlinear} one obtains :

\begin{equation}
C(B,\Delta B,t)=\frac{1}{2}\{[(\rho _{coop}(2B+\Delta B)+\rho
_{diff}(\Delta B))-(\rho _{mc}( B)+\rho _{mc}(B+\Delta
B))]t^{2/3}+S(B,\Delta B)\} \label{correqt}
\end{equation}
 where $S(B,\Delta B)=\ln A(B,\Delta B)-\ln A^{^{\prime
}}(B+\Delta B)-\ln A^{^{\prime }}(\Delta B)$, are  logarithmic
corrections from the
prefactors. $\rho _{mc}(B)$ defines the magneto-conductance, $%
\overline{[J^{*}(B)J(B)]^{n}}=$$\exp \{\sum \frac{n^{i}}{i!}%
C_{i}[lnJ^{*}(B)J(B)]\}=$$A(n,B,\Delta B)\exp (4\ln n+\rho_{mc}
(B)n(n^{2}-1)t)$  \cite{MedinaRev}.

From eqn.\ref{correqt}  apart from the predicted $t^{2/3}$
scaling, one can gain qualitative and quantitative understanding
of the decaying behavior of the correlation function for small
$\Delta B$. With the explicit field dependence of the cooperon and
diffuson given by:
\begin{equation}
\rho_{fluctuations}=(2^{2}/3)(\cos((2B+\Delta
B)/B_{c}^{'})+\cos(\Delta B/B_{c}^{'})+1)\rho(B=0)t^{2/3}
\end{equation}\label{fluctuations}
and
\begin{equation}
\rho_{mc}=(1/3)[(2+\cos(B/B_{c}))+(1/3)(2+\cos((B+\Delta
B)/B_{c})]\rho(B=0)t^{2/3}
\end{equation}\label{magnectoconductance}
one obtains:
\begin{equation}
C(B,\Delta B,t)=\frac{1}{6}{[8\cos((B+\Delta
B)/B_{c}^{'})\cos(B/B_{c}^{'})-2\cos((2B+\Delta
B)/2B_{c})\cos(\Delta B/2B_{c}) ]{\rm Var}_{d}[F(0,t]t^{2/3}}
\end{equation}\label{correqt2}
Here, we have ignored the logarithmic corrections from
eqn.\ref{correqt}.
 For $B=0$ and $\Delta B=0$ one has
$C(B=0,\Delta B=0,t)={\rm Var}_{d}[F(0,t]$, with ${\rm
Var}_{d}[F(0,t]=\rho(B=0)t^{2/3}$, in accordance with
eqn.\ref{nonlinear}
 . Now, for small $\Delta B$
and $B\approx0$, one can assume that $B_{c}^{'}\cong
B_{c}$\cite{MedinaRev}:
\begin{equation}
C(B\approx 0,\Delta B,t)=[1-\frac{1}{2}(\Delta B/B_{c})^{2}]{\rm
Var}_{d}[F(0,t]t^{2/3} \label{correqt3}
\end{equation}
This behavior is is qualitatively seen even for $B\neq0$ in
fig.\ref{fig3}, where for given $B$ one observes a faster decay
with increasing $t$( recall that $B_{c}=\pi/t^{3/2}[\phi_{o}]$).
This is even more clearly to see with SO. In that case one has:
\begin{equation}
 C_{spinor}(B,\Delta
B,t)=\frac{1}{2}[(\rho_{diff} ^{spinor}(\Delta B)-2\rho
_{mc}^{spinor}]t^{2/3}\label{correqtso}
\end{equation}

\begin{equation}
 C_{spinor}(B,\Delta
B,t)=\frac{1}{2}([(2^{2}/2)(\cos(\Delta
B/B_{c}^{'})+1)]-2)\rho_{mc}^{spinor}t^{2/3}\label{correqtso1}
\end{equation}
again an expansion in small $\Delta B$ :
\begin{equation}
 C_{spinor}(B,\Delta
B,t)=(1-\frac{1}{2}(\Delta B/B_{c})^{2}
)\rho_{mc}^{spinor}t^{2/3}\label{correqtso2}
\end{equation}

 On the other hand,
without SO, numerical calculations (fig.\ref{fig3}), shows a very
slow decaying, such that for the field range  less than $\phi
_{o}/\ell^{2}$, there are no enough ensembles to form, such that
even if the field correlation function decays, and a decorrelation
field can be defined, it is not possible to define a large enough
number of samples such that averaging over the field, could be
equivalent to averaging over disorder (samples), a matter that has
caused some confusion. With SO the decaying of the correlation
function is dominated by an even  much bigger field scale unknown
to us.
\smallskip

 Although the replica results are valid quantitatively for $1+1$
dimensions, nevertheless, $2+1$ dimensions also has a bound state, although
weaker, with a smaller positive MC. Therefore, our results apply
qualitatively to three dimensional hops\cite{RRangel2}.

\section{Discussion  and Conclusions}
\label{sec:conclusion}

 Summarizing the results of this work, we
have found that the  log-conductance in the DIR of mesoscopic
samples is  non-ergodic in the mean square sense and the
fluctuations of the log-conductance are non-ergodic in the sense
that sample to sample fluctuations are larger  than magnetic field
fluctuations. Without SO due to the ever increasing
magneto-conductance with the field\cite{MedinaRev,RRangel1}, only
quasi stationarity can be achieved for fields larger than a
certain hopp dependent field $B_{s}(t)$, whereas with SO due to
the small magneto-conductance and its rapid
saturation\cite{MedinaRev,RRangel2}, $B_{s}(t)$ is negligible, so
that the process in that case can be regarded as stationary
independent of t. The field $B_{s}(t)$ is obtained  numerically by
varying $B_{i}$ until one observes that  $ \sigma
_{mss}(B_{f})\approx{\rm Var}_{d}(\langle F({\cal H},B)\rangle)$ .
This different behavior with and without SO turns out to have a
remarkable influence on the behavior of the two variances that
define the sample to sample or disorder fluctuations and  the
field fluctuations, as their values depend on $B_{i}$. It is only
when the averaging interval is taken as $\Delta B=[
B_{s}(t),B_{f}]$,i.e., when quasi-stationarity is achieved, that
the proper comparison of the variance can be made. In experiments
the hopp values were small, and therefore $B_{s}(t)$, was small.
However, in order to compare the magnitude of the fluctuations
defined by eqn.\ref{vardis} and eqn.\ref{varfield} with
experiments one must be aware that their values are sensitive to
[$B_{s}(t)$,$B_{f}$], an important point indicated by our results.
The experiments of \cite{Koslov} were done without SO; therefore,
in order  to make a a rigorous comparison  with our results we
have to know the experimentally taken interval
[$B_{s}(t)$,$B_{f}$] for the evaluation of this quantities. For
the small values of t, according to our results, the influence  of
$B_{i}$ is small  and one should expect $B_{s}(t)\approx0$ and
indeed  we find the ratio of eqn.\ref{vardis} and
eqn.\ref{varfield} predicts a value similar to their experiments.
On the other hand, the scaling of the disorder fluctuations with
the hopp length t, as shown in fig.4, agrees with the experimental
values obtained by Orlov et tal\cite{OrlovRaikhRuzinSav}(see
fig.45 in the review of Kramer and Mackinnon). It is interesting
to observe  that both the variance with disorder and the one with
field appears to have the same functional form.
 A more careful comparison can be done by taking
into account all predictions of this work. In the case with SO,
experimental studies of fluctuations are lacking. To our knowledge
the only work where field fluctuations  in the strongly insulating
regime have been observed in samples with SO  is the work of
Hernandez and Sanquer \cite{HernandezSanquer}. Here  our
predictions give a clear qualitative way of differentiating the
cases with and without SO and highlight a way of evaluation of
experimental data.

From the theoretical point of view, we have derived two important
objects: the ${\it cooperon}$ and the ${\it diffuson}$ which  are
the weak localization analogs in the DIR. They allow us to explain
some features of the correlation function like the qualitative
behavior for small $\Delta B$, and numerically the weak decaying
behavior on the scale $(\phi_{o}/\ell^{2})$, that according to the
Slutski's theorem, is  responsible for the non-ergodic behavior as
we have found.
 In order to experimentally observe persistent diffuson
fluctuations, one has to explore a range of parameters so that
there is a saturation in the average behavior while the wave
function shrinkage\cite{EfrosShklovskii} is still a negligible
effect. This range can be defined by the condition $B_{c}<\hbar
/(ea_{B}N^{1/3})=B_{orb}$, where $a_{B}$ is the Bohr radius.
$B_{orb}$ is
the scale for the orbital shrinkage to be important\cite{LadieuMaillySanquer}%
, i.e., when the cyclotron radius becomes of the order of the mean
free path $\ell $. These conditions have been met in ref.
\cite{Koslov} and \cite {LadieuMaillySanquer}. Furthermore,
according to references\cite
{LadieuMaillySanquer,Koslov,LadieuBouchaud} the magnetic field
cannot induce geometric fluctuations due to changes in the
identity of the hop\cite{Lee}. This finding  holds   in  both two
and three dimensions,  and therefore we expect  the insulating
cooperon and diffuson fluctuations should be seen experimentally
also in three dimensions. In experiments , one should be aware
about the condition of mesoscopic sample,  discussed in section
\ref{sec:NSSmodel}. Otherwise, trivial self-averaging of spatially
different oriented hops can wash out the possibility to observe
fluctuations.

 There will be a
non-ergodic to ergodic behavior in the case when  one relaxes the
condition of DIR with ${\ell }<<\xi <t$ such that there are many
impurities within $\xi $. In this case there will be a diffusing
behavior within the length scale $\xi $ so that two overlapping
random processes  are at work, one ergodic for the diffusing scale
$\xi$ and the other  non-ergodic for the larger scale $t$. This
question is under study \cite{FengPichard}. Finally, we want to
stress that the fundamental point of considering correlations in
the random processes $F({\cal H},t)$ due to the crossing of paths
can not be relaxed. This approach permitted us to use the full
power of the replica theory, which again helped to
semi-quantitative and qualitative predictions. Theories like the
independent path approximation, where such crossing is  neglected,
miss the very crucial objects defined by the cooperon and the
diffusson.


\section{Acknowledgments}
One of us (RR) wants to express his gratitude to the  head of
department  of ICA1, Prof. Hans J. Herrmann, for the hospitality
and support. Also (RR) one to express  his gratitude to Dr.
Maheboob Alam for helping me in technical details with xmgr and
making helpful corrections in the written english. Author
correspondence email: $\diamond$rerangel@usb.ve
 \label{sec:agradecimientos}

\newpage

\begin{center}
\parbox{0.90\textwidth}{\epsfig{file=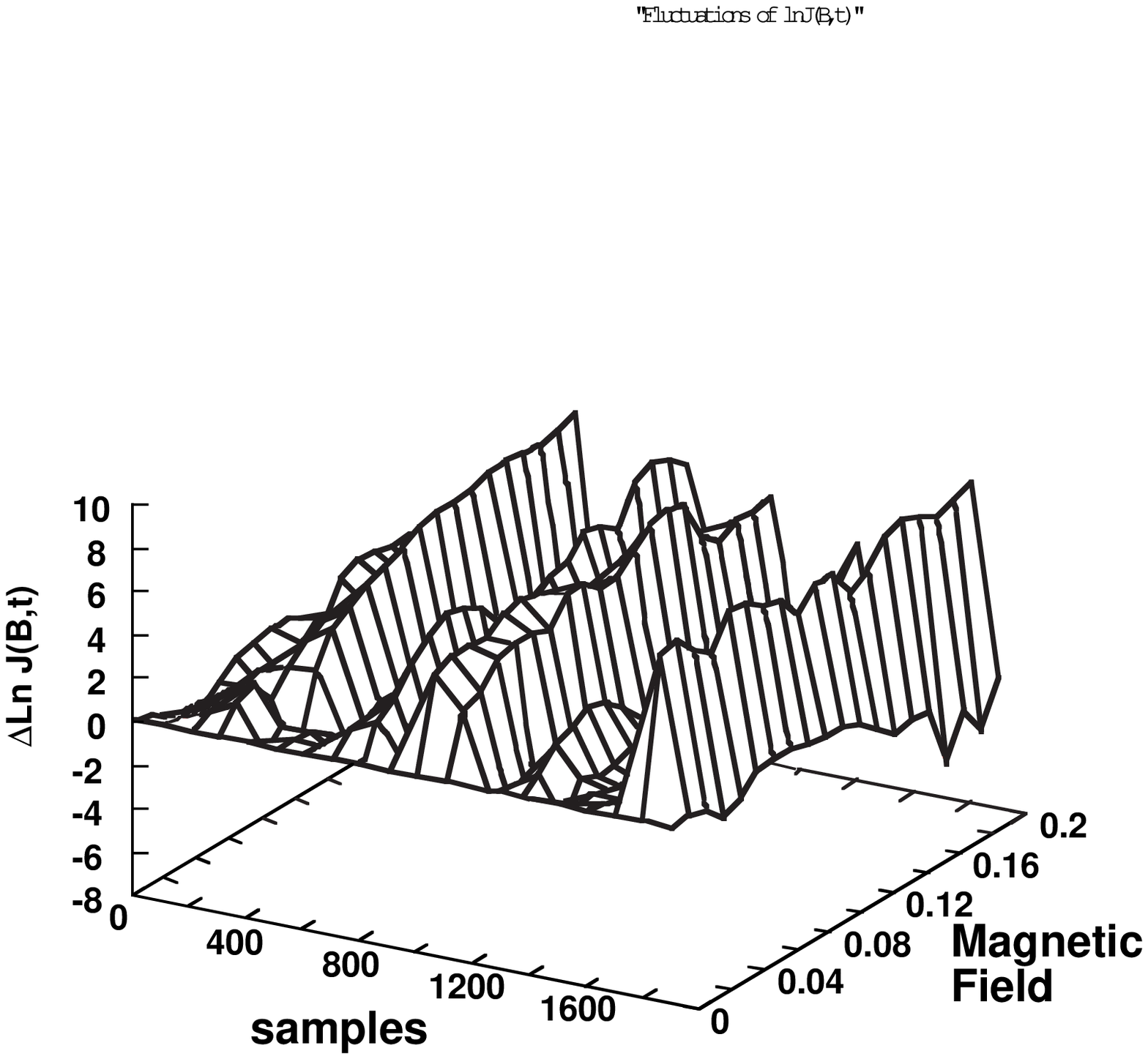,width=0.90\textwidth,angle=-00}}
\end{center}
\newpage
\begin{figure}[tbp]
\caption{Field Conductance fluctuations of $ln|J(B,t)|$ and
$ln|I(B,t)|$  for different realizations of disorder  or sample
number and   $t=15$ } \label{fig1}
\end{figure}
\newpage

\begin{center}
\parbox{0.50\textwidth}{\epsfig{file=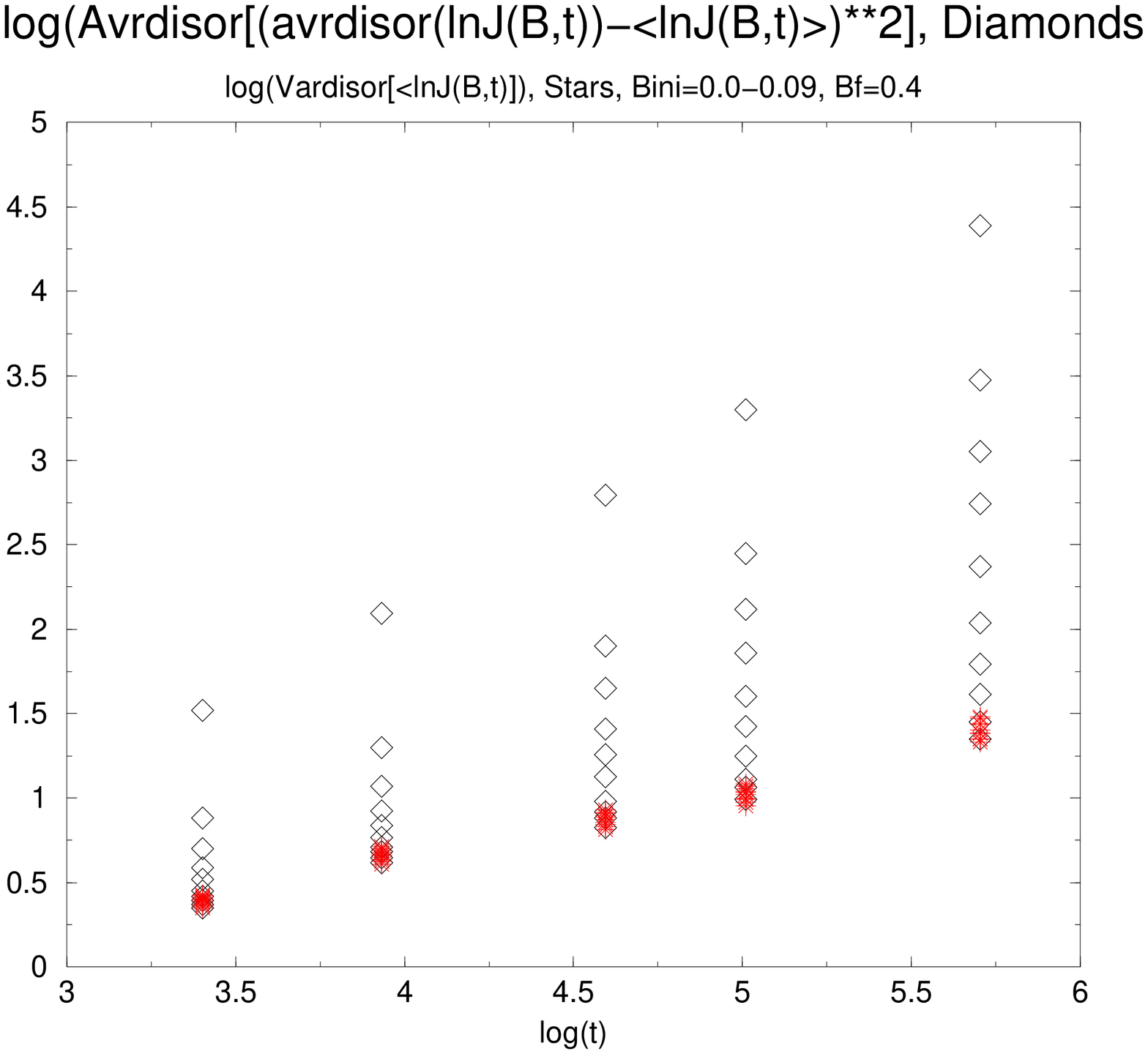,width=0.50\textwidth,angle=-90}}
\end{center}
\begin{center}
\parbox{0.50\textwidth}{\epsfig{file=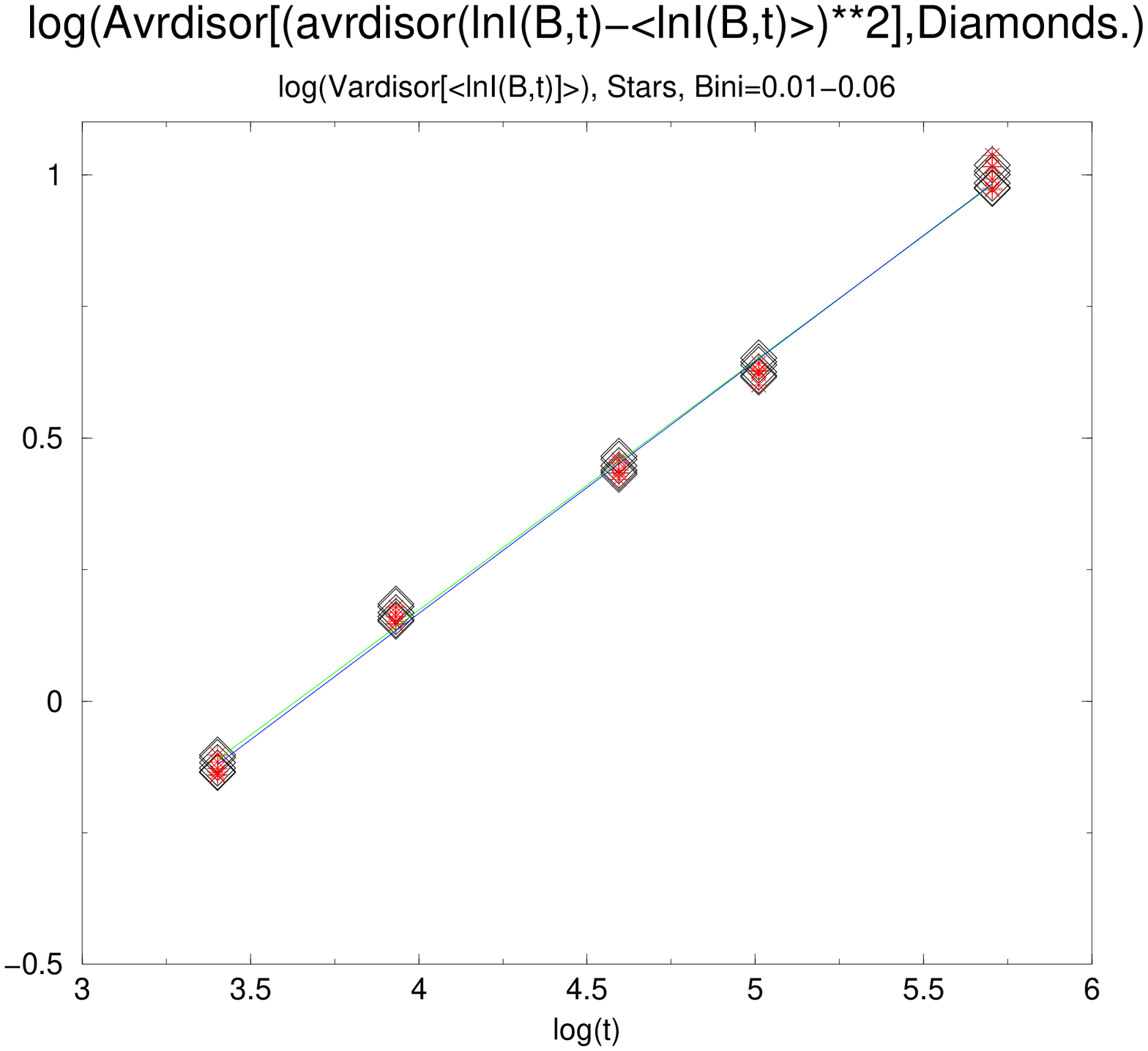,width=0.50\textwidth,angle=-90}}
\end{center}

\begin{figure}[tbp]
\caption{ Quasi Stationary behavior without SO. Determining
$B_{s}(t)$ for t=30,50,100,150,300. Diamonds converge to stars
symbols as the field $B_{i}$ grows from zero  to 0.09 in 0.01
steps. In the case with  SO diamonds and stars fall together
independent of $B_{i}$ and $t$ indicating stationarity with
$B_{s}\approx{0}$ } \label{fig2}
\end{figure}

\newpage

\begin{center}
\parbox{0.50\textwidth}{\epsfig{file=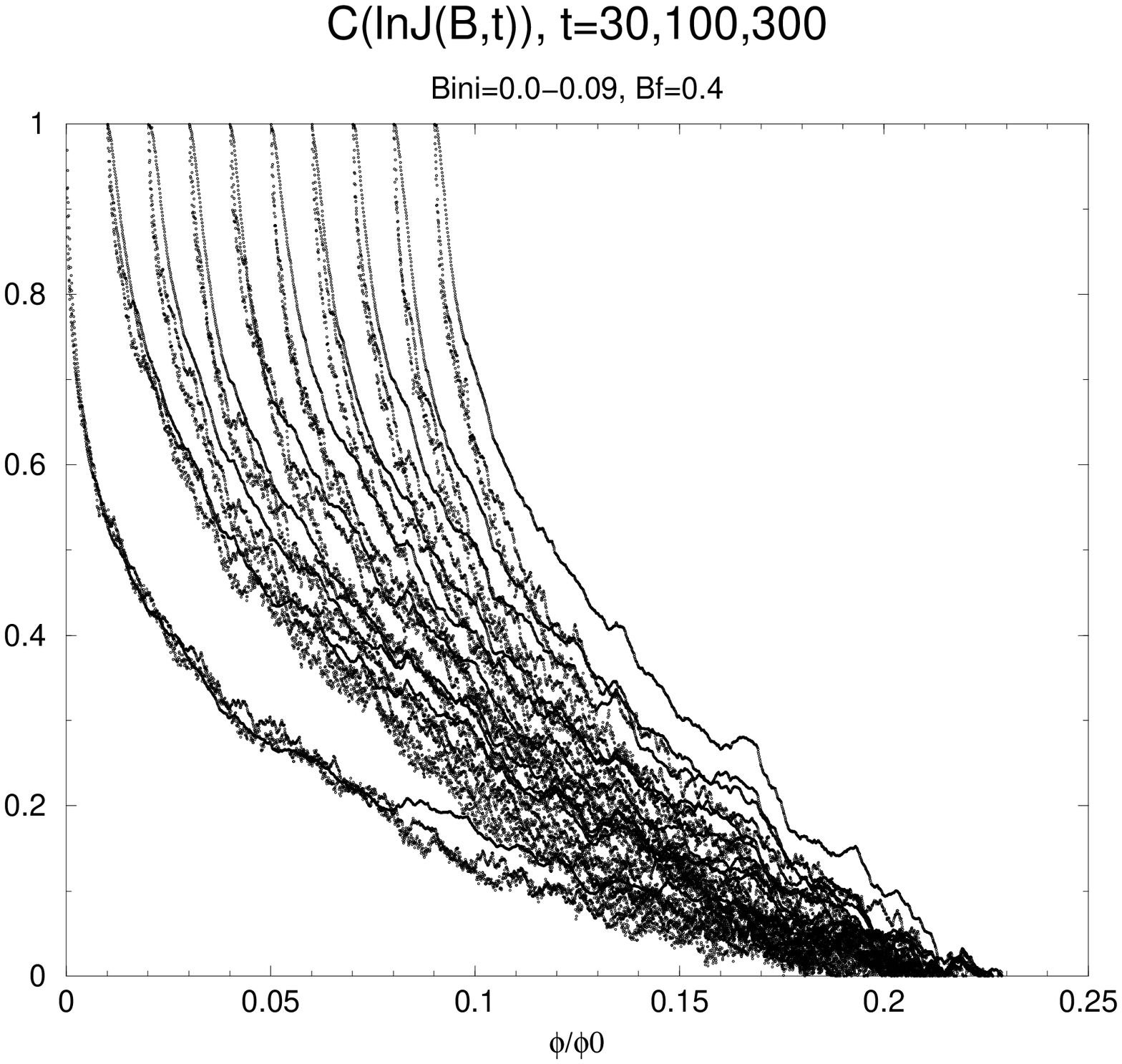,width=0.50\textwidth,angle=-90}}
\end{center}
\begin{figure}[tbp]
\caption{Correlations functions $C(B,\Delta B,t)$  and
$C_{spinor}(B,\Delta B,t)$ as a function of $\Delta
B$($\phi/\phi_{o}$) for a given value of $B$ from $B=0.0$ to
$B=0.09$. For each value of $B$, there are three curves
corresponding to three values of $t$$(30,100,300)$. One sees a
tendency to faster decaying with increasing $t$ .}
 \label{fig3}
\end{figure}

\newpage
\begin{center}
\parbox{0.50\textwidth}{\epsfig{file=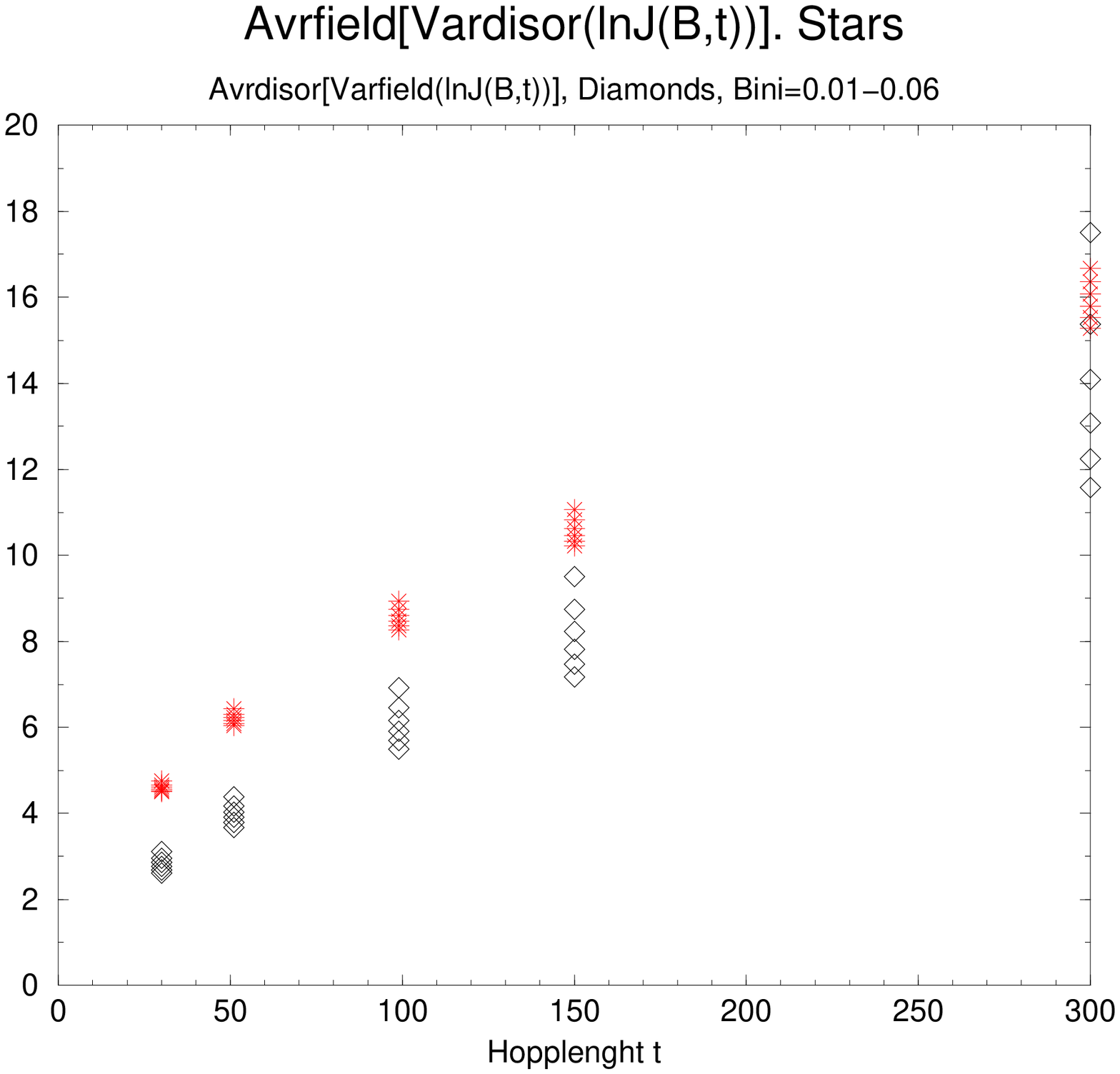,width=0.50\textwidth,angle=-90}}
\end{center}
\begin{center}
\parbox{0.50\textwidth}{\epsfig{file=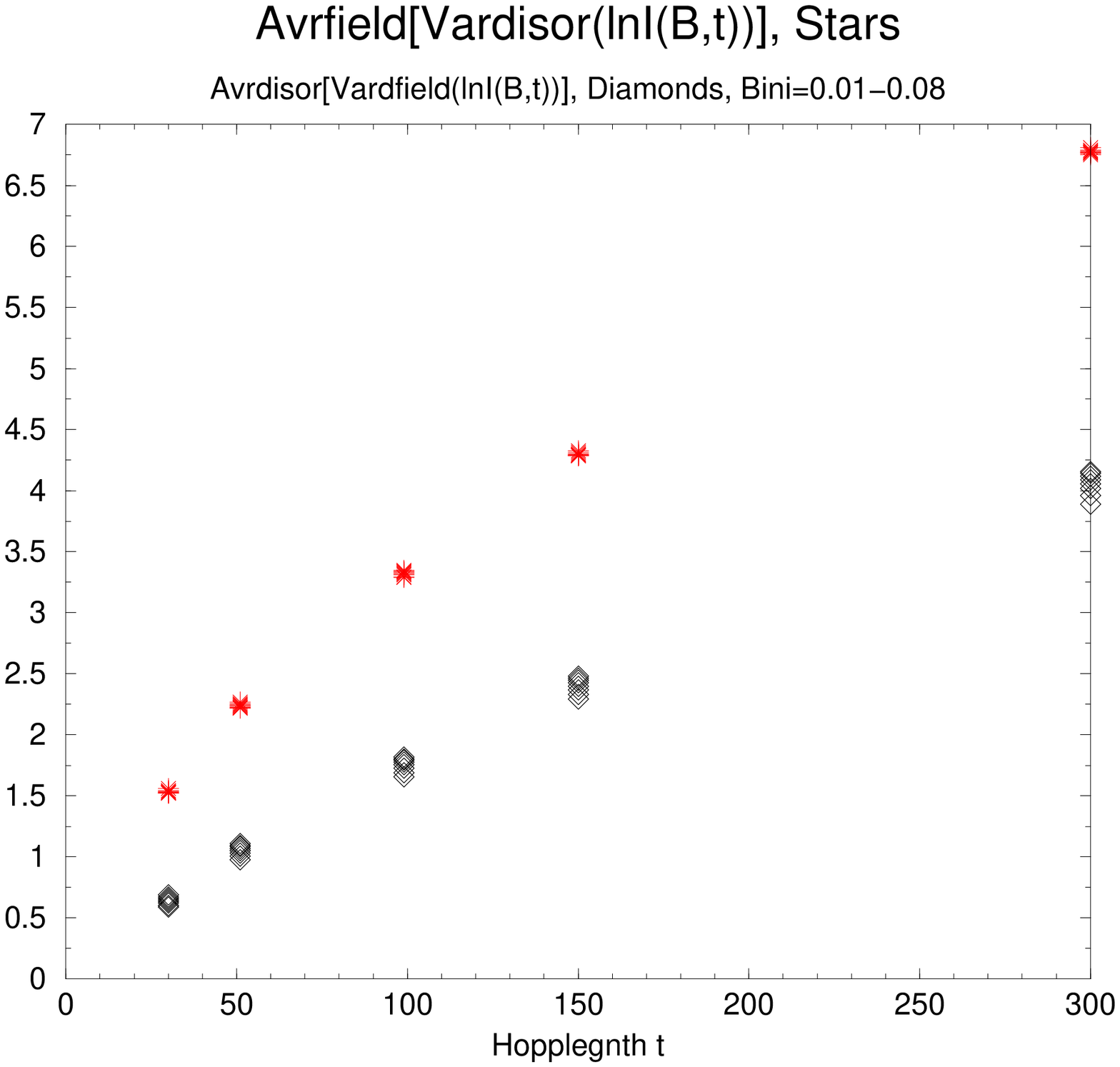,width=0.50\textwidth,angle=-90}}
\end{center}
\begin{figure}[tbp]
\caption{The figure depicts the variance as defined by
Eq.\ref{vardis} and Eq.\ref{varfield} as a function of t. The
effect of changing  the  values of $B_{i}$ from $0$ to $0.08$ is
clearly seen until   saturation is attained. Notice that there is
a crossing due to a stronger with $t$ increasing  dependence on
$B_{i}(t)$ on eqn.\ref{vardis} than in eqn.\ref{varfield}. In the
case with SO, both quantities have a weak dependence on $B_{i}$
and the crossing efect is absent.}
 \label{fig4}
\end{figure}

\newpage
\begin{center}
\parbox{0.50\textwidth}{\epsfig{file=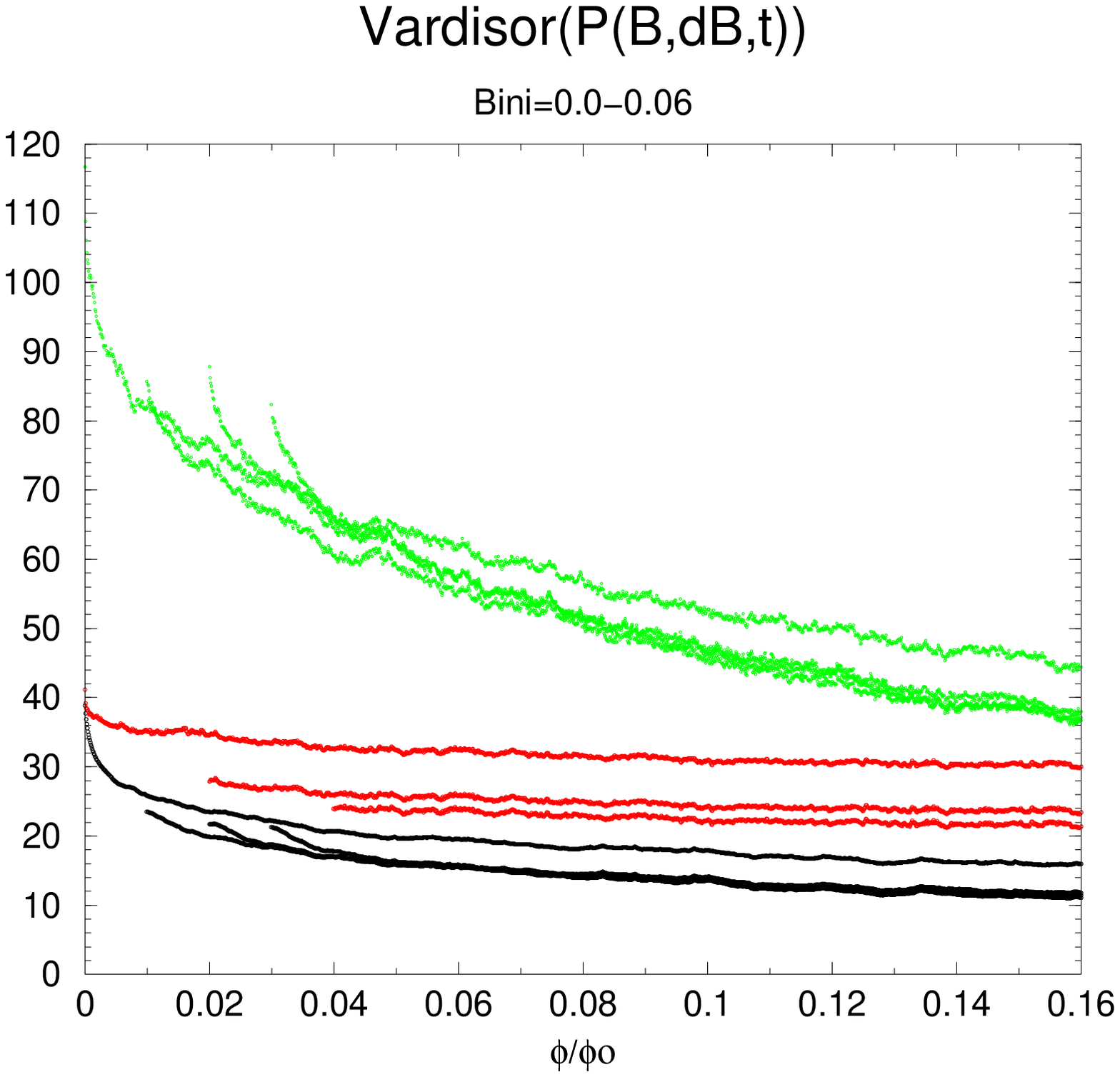,width=0.50\textwidth,angle=-00}}
\end{center}
\begin{center}
\parbox{0.50\textwidth}{\epsfig{file=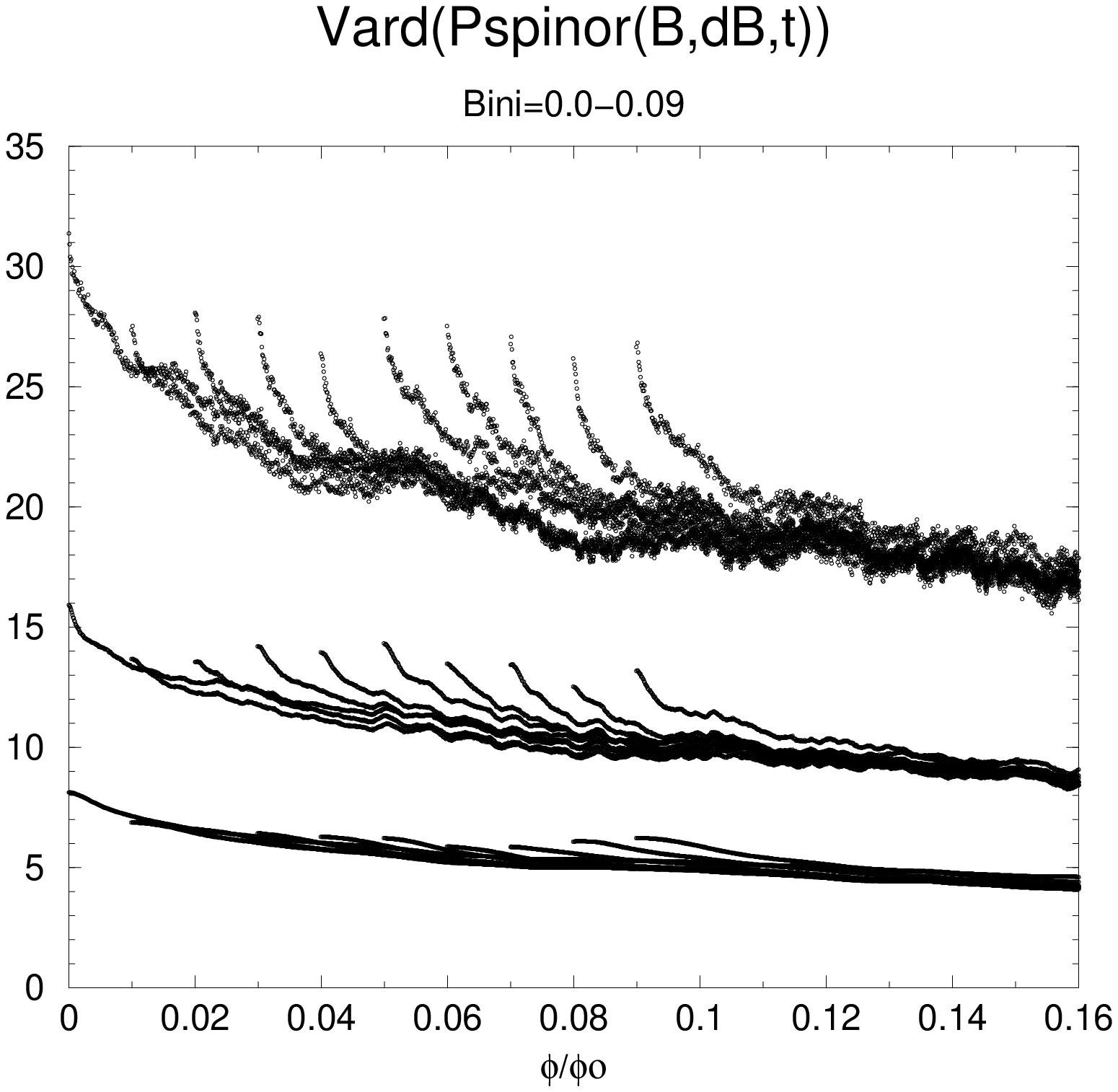,width=0.50\textwidth,angle=-00}}
\end{center}
\begin{figure}[tbp]
\caption{The variance in disorder of $P(B,dB,t)$ and
$P_{spinor}(B,dB,t)$ as a function of $dB$($\phi/\phi_{o}$) for
$t=30,100,300$ and $B=B_{i}=0.0-0.09$.}
 \label{fig5}
\end{figure}

\newpage

\begin{center}
\parbox{0.50\textwidth}{\epsfig{file=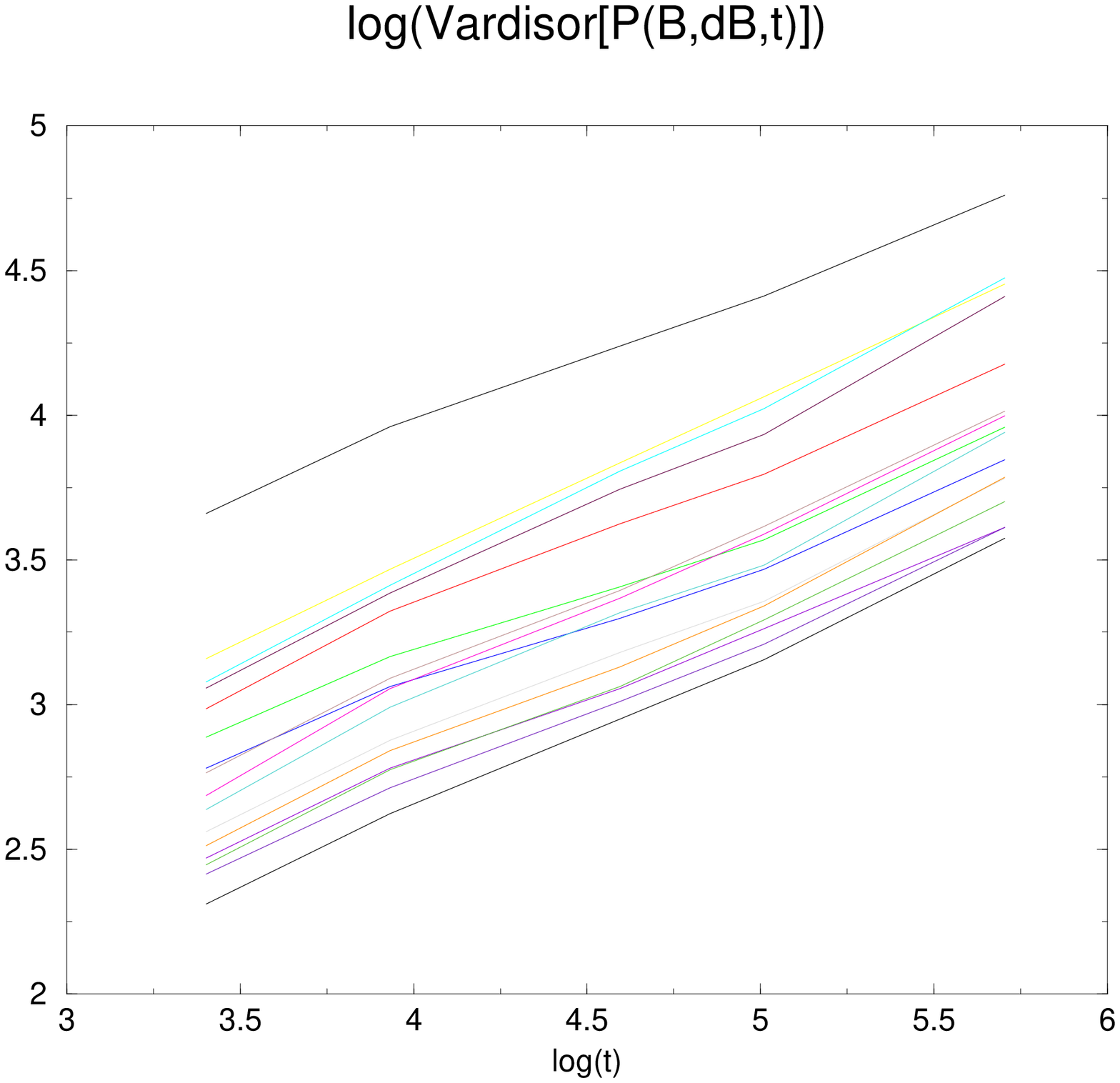,width=0.50\textwidth,angle=-90}}
\end{center}
\begin{center}
\parbox{0.50\textwidth}{\epsfig{file=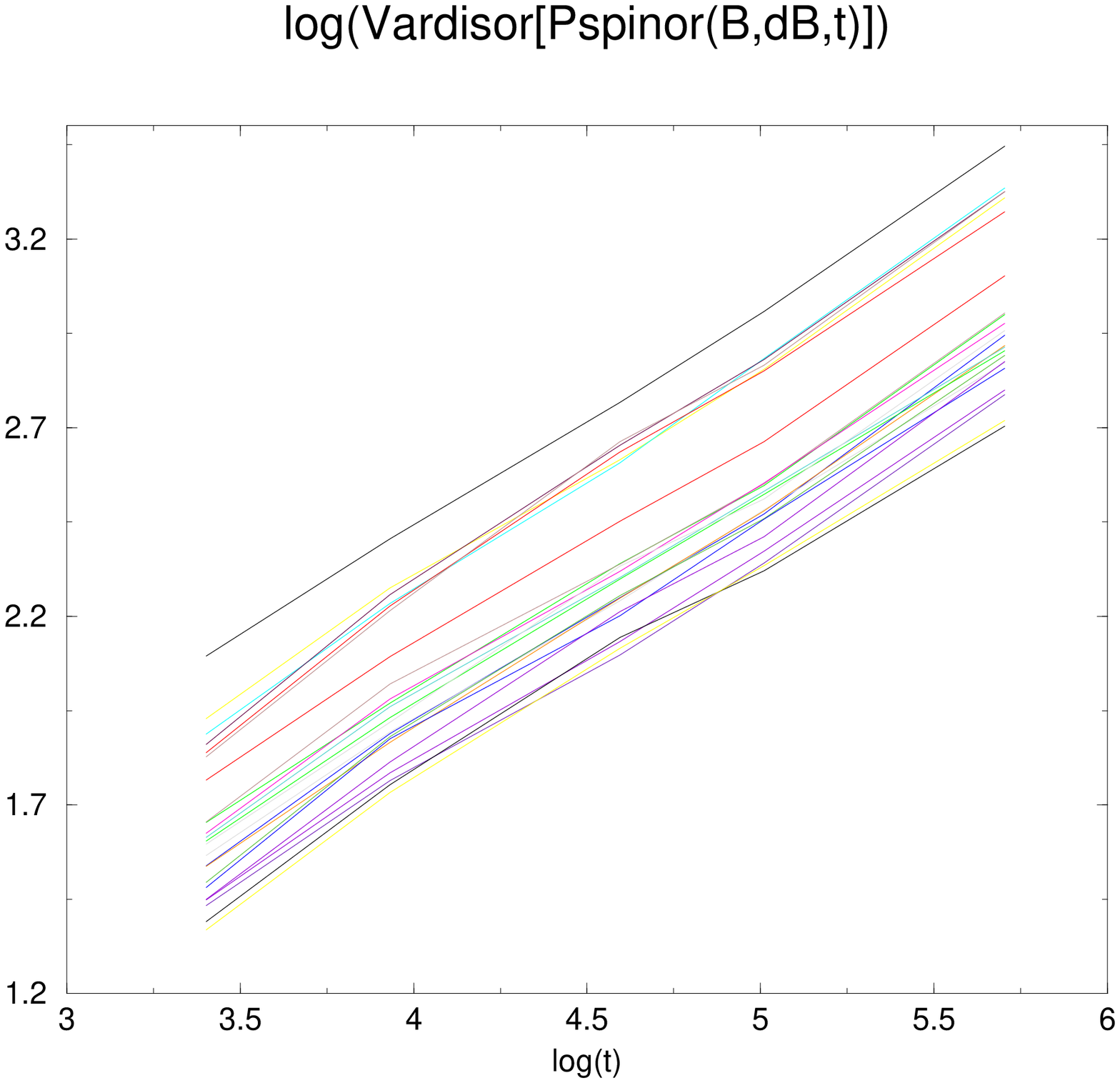,width=0.50\textwidth,angle=-90}}
\end{center}
\begin{figure}[tbp]

\caption{The figure shows the expected $t^{2/3}$ dependence from
eqn.\ref{cooperondiff}. Each line corresponds to a different
${\Delta B}$. The upper line corresponds to ${\Delta B=0}$. With
increasing ${\Delta B}$ the lines move downward and saturate for
big enough ${\Delta B}$. One observes small modulations which are
interpreted as deriving from the normalization factors in the
Bethe Anzatz. }
 \label{fig6}
\end{figure}



\newpage

\begin{center}
\parbox{0.80\textwidth}{\epsfig{file=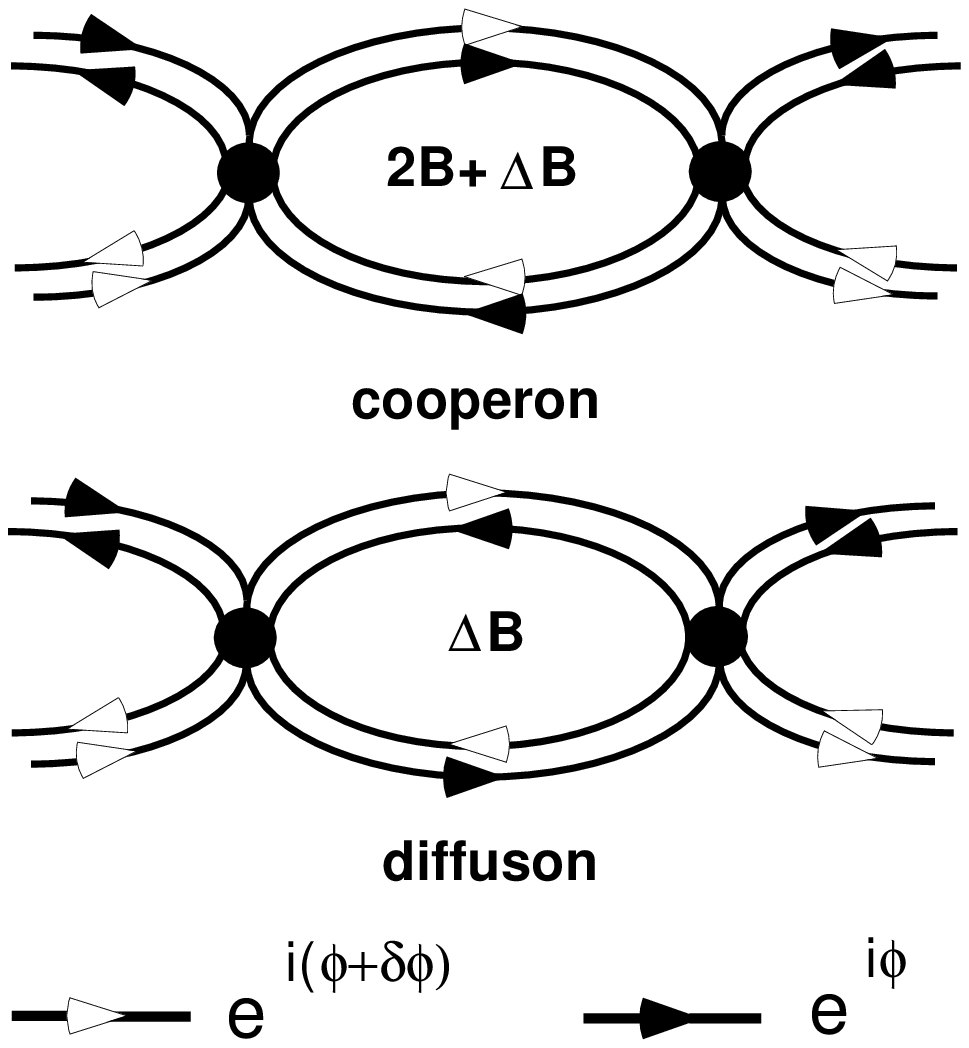,width=0.80\textwidth,angle=-00}}
\end{center}
\begin{figure}[tbp]
\caption{The Cooperon and Diffuson  without SO}
 \label{fig7a}
\end{figure}

\newpage

\begin{center}
\parbox{0.50\textwidth}{\epsfig{file=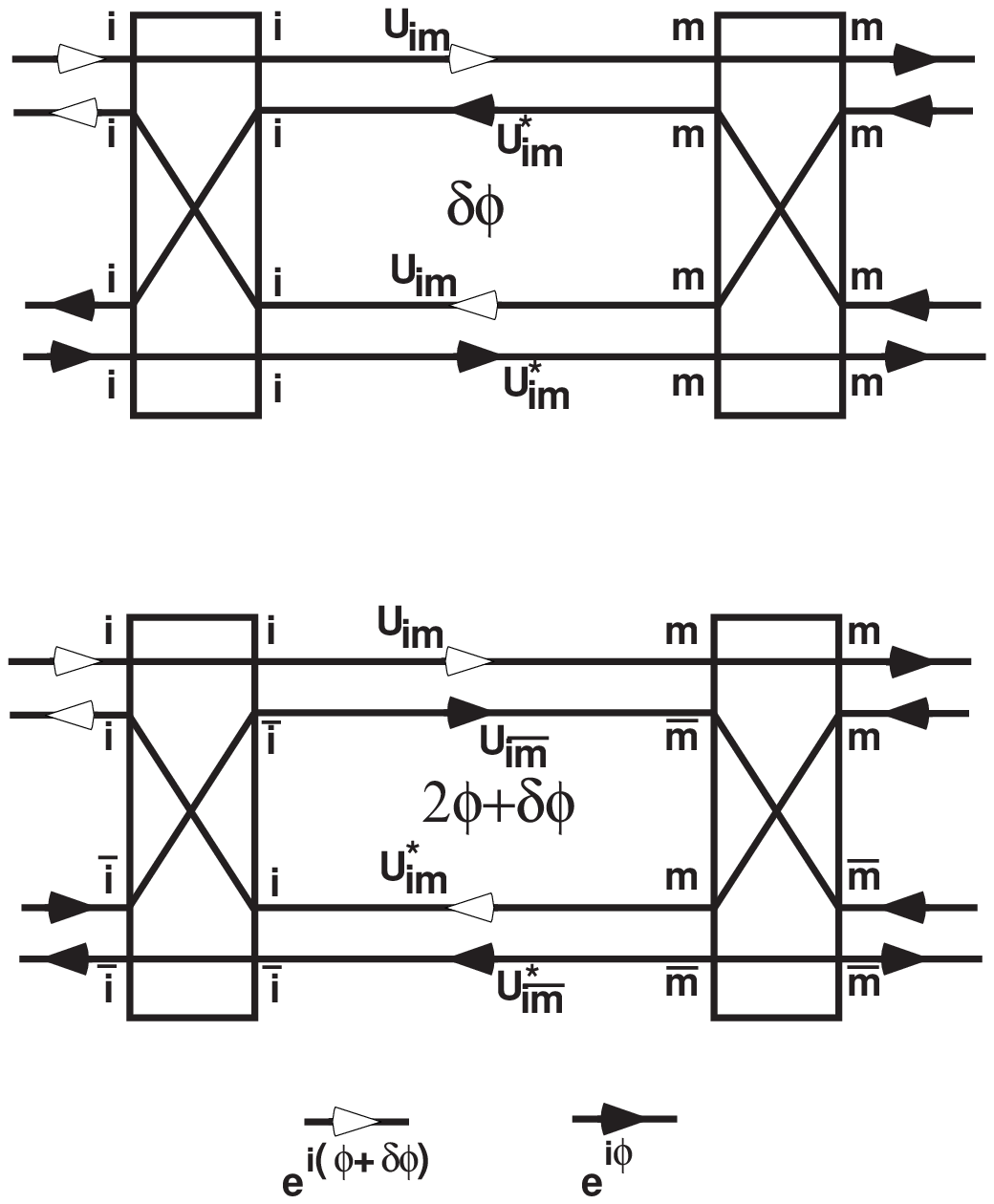,width=0.50\textwidth,angle=-00}}
\end{center}
\begin{figure}[tbp]
\caption{The  Diffuson  and the Cooperon with SO}
 \label{fig7b}
\end{figure}


\begin{thebibliography}{40}
\bibitem{explanationnonlinear}  {This relation expresses a
non-linear property, because  taking the variance of the sum
$[F(B+\Delta B,t)+F(B,t)]$ doest not equal to the sum of the
variance of each term  in the sum, but there is a non-linear
correction given by twice the correlation function.}
\end{thebibliography}
\end{document}